\documentclass[3p,times,procedia]{elsarticle}
\usepackage{nupha_ecrc}
\usepackage{tabularx}
\usepackage{xcolor}
\usepackage{amsmath}

\volume{00}

\firstpage{1}

\journalname{Nuclear Physics A}

\runauth{A. Motornenko et al.}

\jid{nupha}

\jnltitlelogo{Nuclear Physics A}

\usepackage{amssymb}

\usepackage[figuresright]{rotating}

\begin{document}

\begin{frontmatter}



\dochead{XXVIIth International Conference on Ultrarelativistic Nucleus-Nucleus Collisions\\ (Quark Matter 2018)}

\title{QCD at high density:\\ 
Equation of state for nuclear collisions and neutron stars}


\author[ITP,FIAS]{Anton~Motornenko}
\author[ITP,FIAS]{Volodymyr~Vovchenko}
\author[FIAS]{Jan~Steinheimer}
\author[FIAS]{Stefan~Schramm}
\author[ITP,FIAS,GSI]{Horst~Stoecker}
\address[ITP]{Institut f\"ur Theoretische Physik,
Goethe Universit\"at Frankfurt, D-60438 Frankfurt am Main, Germany}
\address[FIAS]{Frankfurt Institute for Advanced Studies, Giersch Science Center, D-60438 Frankfurt am Main, Germany}
\address[GSI]{GSI Helmholtzzentrum f\"ur Schwerionenforschung GmbH, D-64291 Darmstadt, Germany}

\begin{abstract}
A unified chiral mean field approach is presented for QCD thermodynamics in a wide range of temperatures and densities. The model simultaneously gives a satisfactory description of lattice QCD thermodynamics and fulfills nuclear matter and astrophysical constraints.
The resulting equation of state can be incorporated in relativistic fluid-dynamical simulations of heavy-ion collisions and neutron stars mergers.
Access to different regions of the QCD phase diagram can be obtained in simulations of heavy-ion data and observations of neutron star mergers.
\end{abstract}


\end{frontmatter}



\section{Introduction}
\label{sec:introduction}

Current first-principle calculations of the finite temperature QCD equation of state are limited to small values of the baryochemical potential, where lattice QCD methods are applicable~\cite{Borsanyi:2013bia, Bazavov:2014pvz, DElia:2016jqh, Bazavov:2017dus, Borsanyi:2018grb}.
The experimental studies of collisions of heavy ions provide only a limited information on QCD thermodynamics due to
the finite lifetime and size of the system created.

Lattice QCD data at vanishing baryochemical potential $\mu_B=0$ suggest a smooth crossover at $T \approx m_{\pi}$, the chiral transition, where (approximate) chiral symmetry is restored.
Does this crossover turn into a first-order transition at finite baryon number density? What is the relation of the chiral symmetry restoration to the deconfinement?
An analysis of lattice data at zero and imaginary chemical potential at $T\geq 130$ MeV shows no indications for the presence of a critical point in the real $T-\mu_B$ plane at moderate baryochemical potentials~$\mu_B/T \lesssim \pi$~\cite{Vovchenko:2017gkg}. 
For larger chemical potentials, $\mu_B/T>\pi$, one has to rely on effective model descriptions.
These models are usually constrained either by lattice data at zero chemical potential or by nuclear matter properties at zero temperature.

Here a new model is introduced, where both constraints, at zero $\mu_B$ and at zero $T$ are used successfully.

\section{The CMF model}

The SU(3) flavor parity doublet quark-hadron chiral mean field model (CMF model) \cite{Steinheimer:2010ib,Steinheimer:2011ea,Mukherjee:2016nhb} is a unified description of the statistical and thermodynamical properties of QCD matter. The complete description of QCD bulk properties includes a complete list of known hadrons (with masses below 2.6 GeV) as well as the three light quark flavors. The transitions between quark matter and a hadron-resonance gas (HRG) as well as nuclear matter are driven by mean fields. Nucleons and other baryons of the respective SU(3) flavor representation (octet and decuplet) interact via mesonic mean fields ($\sigma,~\omega,~\rho,~\phi,~\zeta$) in a non-linear $\sigma-\omega$ model approach. Properties of nuclear matter are reproduced \cite{Papazoglou:1998vr,Dexheimer:2007tn}. Parity doubling introduces heavy parity partners to the baryons of the lowest flavor multiplets \cite{Steinheimer:2011ea}. Hence an explicit mass term for baryons in the Lagrangian is possible, which preserves chiral symmetry. The effective masses of the parity partners depend on the chiral fields, therefore the partners become mass-degenerate as chiral symmetry is restored:
\begin{eqnarray}
m^*_{i} = \sqrt{ \left[ (g^{(1)}_{\sigma i} \sigma + g^{(1)}_{\zeta i}  \zeta )^2 + (m_0+n_s m_s)^2 \right]} \pm g^{(2)}_{\sigma i} \sigma \pm g^{(2)}_{\zeta i} \zeta ~,
\label{eq:bar_mass}
\end{eqnarray}
This approach is supported by recent lattice calculations, which do indeed show that the masses of the parity partners approach the same value above the pseudocritical temperature~\cite{1710.08294}.

The quarks are treated here as in the PNJL-like approach \cite{hep-ph/0310121}. The effective quark mass $m_{q}^*$ is dynamically generated by the chiral fields $\sigma$ and $\zeta$ (non-strange and strange quark condensates). The quark contribution to thermodynamic potential $\Omega_{q}$ is controlled by the Polyakov loop order parameter $\Phi$, the value of $\Phi$ is determined by the potential $U(\Phi)$ \cite{Ratti:2005jh}:
\begin{gather}
	\Omega_{q}=-T \sum_{i\in Q}{\frac{\gamma_i}{(2 \pi)^3}\int{d^3k \ln\left(1+\Phi \exp{\frac{E_i^*-\mu_i}{T}}\right)}}\,,~
m_{q}^*=-g_{q\sigma}\sigma+\delta m_q + m_{0q}\,, ~ m_{s}^*=-g_{s\zeta}\zeta+\delta m_s + m_{0q}\,,\nonumber\\
	U=-\frac12 (a_0 T^4+a_1 T_0 T^3+a_2 T_0^2 T^2)\Phi\Phi^*
	+ b_3 T_0^4 \log[1-6\Phi\Phi^*+4(\Phi^3 + \Phi^{*3})-3(\Phi\Phi^*)^2]
\label{eq:pnjl}
\end{gather}
The additional mass shift $\delta m_q$ for quarks is motivated by gluon contributions to the effective quark mass. It prevents quarks from appearing in the nuclear ground state.

Excluded volume corrections implemented for hadrons \cite{Rischke:1991ke} prevent a ``reconfinement'' at high densities \cite{Steinheimer:2010ib}:
\begin{eqnarray}
\rho_i=\frac{\rho^{\rm id}_i (T, \mu_i^*)}{1+\sum\limits_j v_j \, \rho^{\rm id}_j(T, \mu_j^*)} \, .
\end{eqnarray}
The excluded volume parameter $v_j$ is set to $v_B = 1$ fm$^3$ for baryons and to $v_M = 1/8$ fm$^3$ for all mesons.

\section{Comparison with the lattice data}
\label{sec:comparison}

First-principle lattice calculations of QCD thermodynamics provide a valuable input to phenomenological models, as a basis for calculations at finite baryon densities.
To constrain the free parameters of the quark sector of the CMF model we use the interaction measure $I$ at $\mu_B=0$ as a representative measure to describe the thermodynamics of the transition from hadron to quark degrees of freedom. We treat the parameters of the Polyakov loop potential and quark couplings to the chiral fields as free parameters. Lattice QCD data on the interaction measure are well reproduced when the following parameters are used:
\begin{center}
$T_0= 180$ MeV, $a_1=-11.67$, $a_2=9.33$, $b_3= -0.53$ and $g_{q\sigma}=g_{s\zeta}=-1.0$\,.
\end{center}

\begin{figure}[h!]
\centering
\includegraphics[width=.32\textwidth,trim={0.1cm 0 3cm 0},clip]{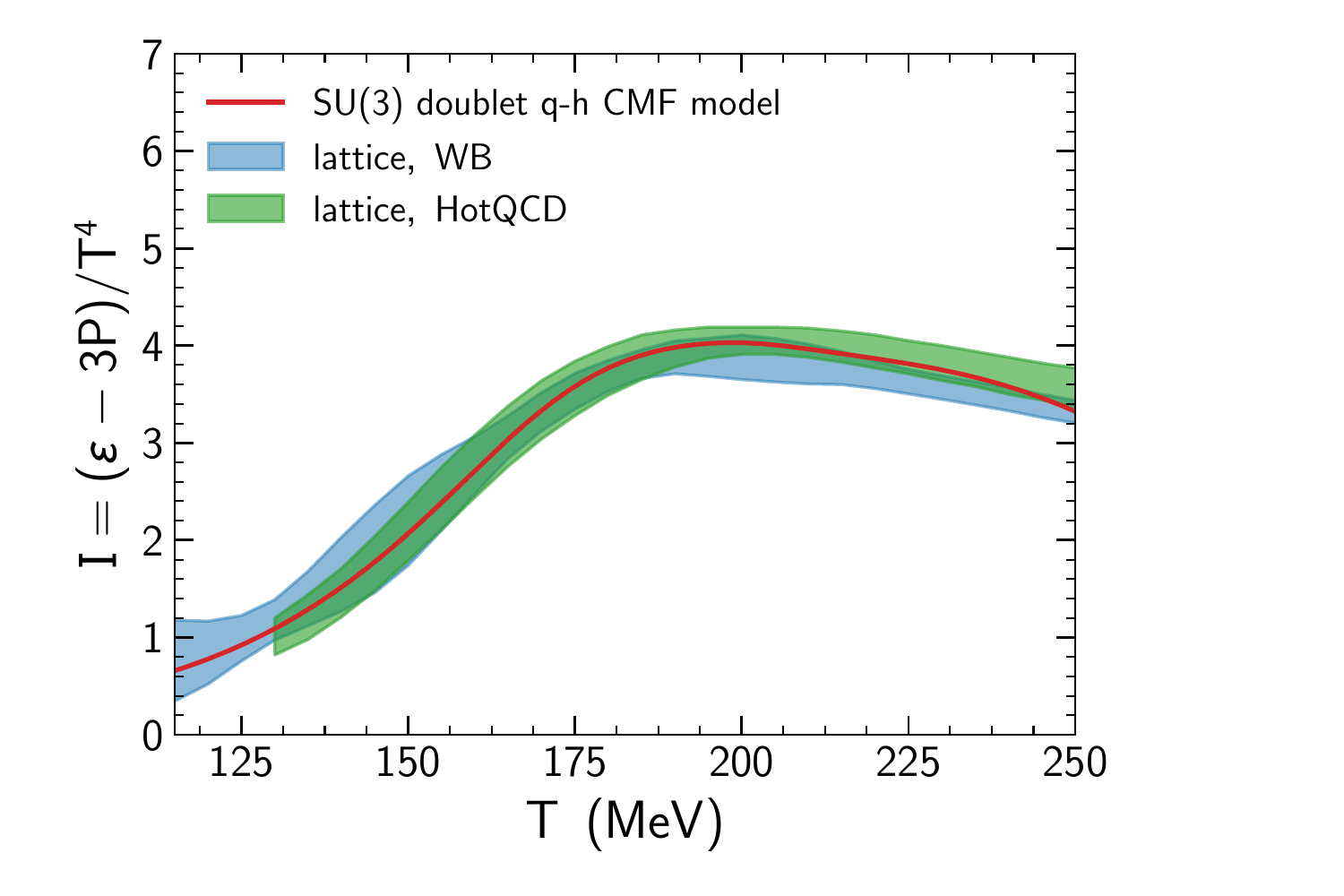}
\includegraphics[width=.32\textwidth,trim={0.1cm 0 3cm 0},clip]{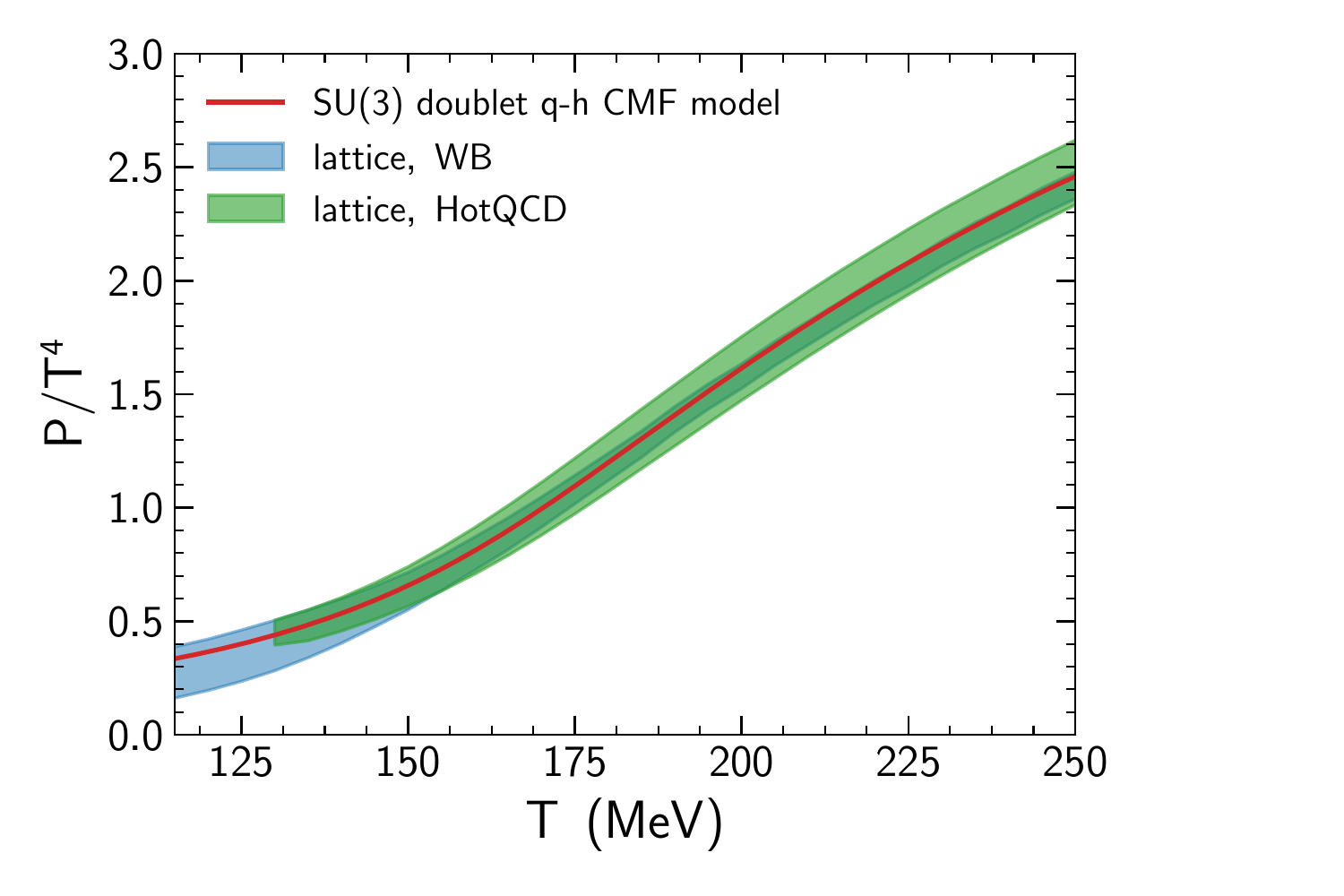}
\includegraphics[width=.32\textwidth,trim={0.1cm 0 3cm 0},clip]{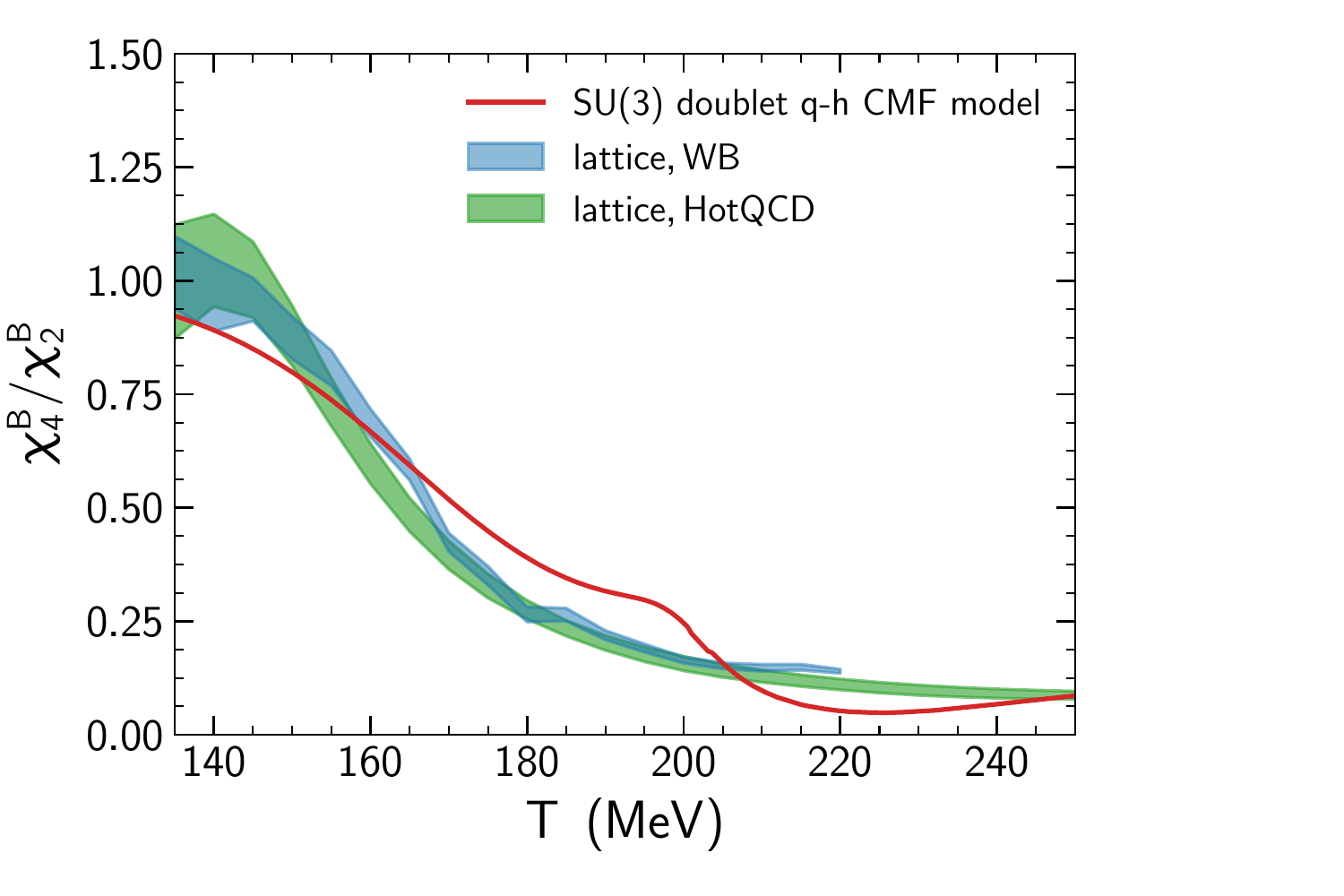}
\caption{Trace anomaly $I$, pressure $P$, and net baryon number susceptibilities ratio $\chi_4^B/\chi_2^B$ at $\mu_B=0$ as a functions of temperature $T$.
The red line depicts the CMF model, blue and green bands show the results of the lattice QCD calculations from Wuppertal-Budapest and the HotQCD collaborations, respectively \cite{Borsanyi:2013bia, Bazavov:2014pvz, Bazavov:2017dus, Borsanyi:2018grb}.}
\label{fig:fit}
\end{figure}

The temperature dependence of the interaction measure and pressure as well as the baryon number susceptibilities 
\begin{eqnarray}
\chi_n^B= \frac{\partial^n (P/T^4)}{(\partial \mu_B/T)^n}\,,
\label{eq:susc}
\end{eqnarray}
are shown in in Fig.~\ref{fig:fit}. 
The lattice data on the kurtosis $\chi_4^B/\chi_2^B$ and the CMF model predictions show a smooth transition from $1$ -- the ideal HRG model value -- to $2/(3 \pi^{2})$ -- the value for the Stefan-Boltzmann limit of massless quarks. 
The small bump at $T\approx 200$ MeV, in the CMF model, is not supported by the lattice data. It results from the chiral symmetry restoration which occurs in the model at this temperature.

\section{Finite density applications}
\label{sec:application}
Experimentally, QCD thermodynamics at large densities can be probed in relativistic heavy-ion collisions, by neutron star properties, and, since GW170817, in binary neutron star mergers~\cite{TheLIGOScientific:2017qsa}. 
The densities created in both event-scenarios are in excess of several times nuclear saturation density. 
The QCD equation of state is the common ingredient for fluid-dynamical simulations which connect all these phenomena~\cite{Hanauske:2017oxo}.
Which regions of the phase diagram can be probed in those different physical situations? 

\begin{figure}[h!]
\centering
\includegraphics[width=.49\textwidth]{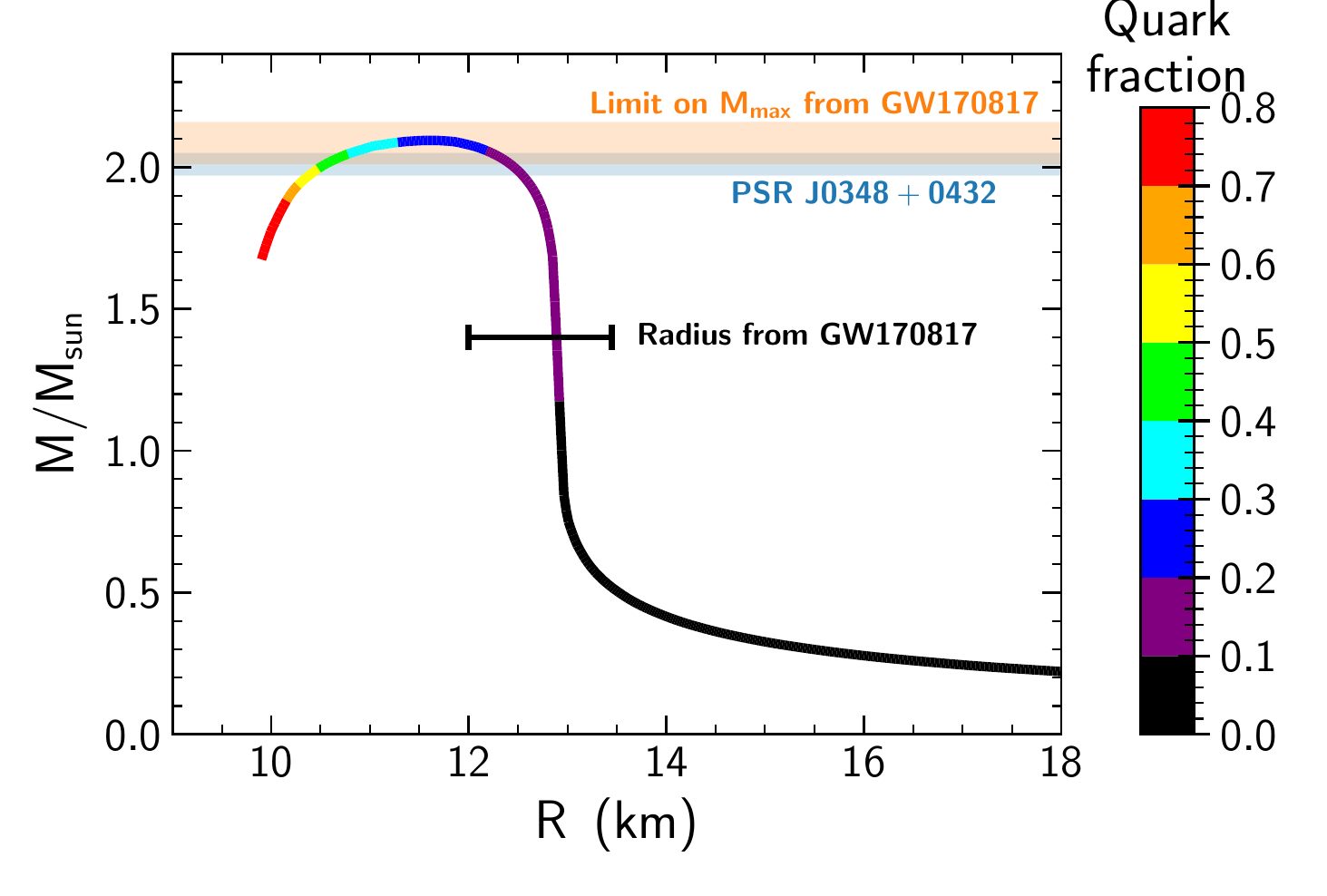}
\includegraphics[width=.49\textwidth]{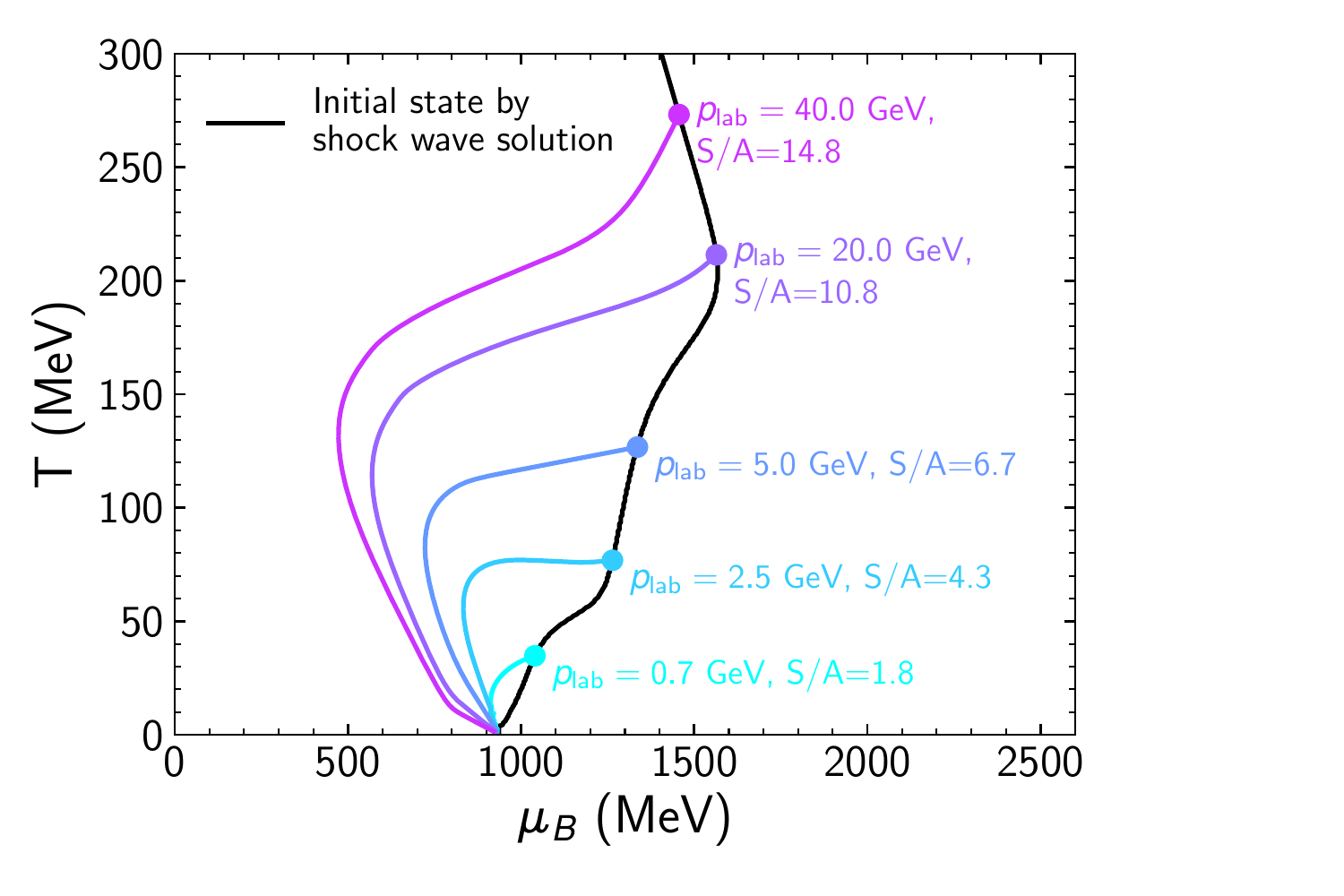}
\caption{Left: Mass-radius diagram of neutron stars in $\beta-$equilibrium and at $T=0$, calculated within the CMF model. Color indicates the fraction of the star's mass due to deconfined quarks. Bands are taken from \cite{1711.00314, 1803.00549}.
Right: The evolution of central heavy-ion collisions through the $T-\mu_B$ plane for different FAIR bombarding beam energies. Black line -- Taub adiabat that describes initial state of heavy ion collisions as implicit function of $\sqrt{s_{\rm NN}}$. Colored lines -- isentropes (lines of constant entropy per baryon $S/A$).}
\label{fig:densities}
\end{figure}

The neutron star mass is related to its radius by the Tolman-Oppenheimer-Volkoff (TOV) equation \cite{Tolman:1939jz,Oppenheimer:1939ne}, which uses the pressure as a function of the energy density at zero temperature as input. 
The astrophysical constraints on neutron star masses and radii then serve as a benchmark for the low-temperature and high-density part of the EoS. The mass-radius relation for compact stars with the CMF model is presented in Fig.~\ref{fig:densities} (left). 
The solution of the TOV equation predicts that stars with a total quark fraction of more than 30\% are unstable. 
The model does not yield a second family of quark stars. There is no first order phase transition for neutron star matter for this CMF model parametrization. 
A rather smooth appearance of the quarks does not lead to a quark core in the center of the star as the quark phase is not spatially separated, by a phase boundary, from the hadrons. 
The predicted CMF mass-radius relation lies within the measured mass and radius constraints for neutron stars \cite{1711.00314, 1803.00549}.

The entropy in heavy ion collisions is mostly produced in the initial state of the collision and, for moderate beam energies, can be estimated using
the Taub adiabate \cite{10.1086/151927} (shock solution). 
As the system expands along a line of constant entropy per baryon, $S/A$, it cools until a dynamical freeze-out range of $T$ and $\mu_B$ values is reached. The predicted isentropes are shown on the right hand side of Fig.~\ref{fig:densities}. At FAIR energies, the system probes temperatures from $10<T<270$ MeV and chemical potentials from $500<\mu_B<1500$ MeV.

In this region the CMF model has one first order phase transition -- the nuclear liquid-gas transition at $T \approx 10$ MeV.
Therefore, there is no critical point associated with a deconfinement transition and the transformation to quarks is smooth.
Signatures of the nuclear liquid-gas transition, on the other hand, appear to be relevant and are measurable and observed even at higher temperatures, probed by the highest energy heavy-ion collisions~(see also~\cite{Mukherjee:2016nhb,Vovchenko:2016rkn,Vovchenko:2017ayq}).

\section{Summary}
\label{sec:summary}
A unified chiral mean field approach is presented for QCD thermodynamics in a wide range of temperatures and densities. The model simultaneously gives a satisfactory description of lattice QCD thermodynamics and fulfills nuclear matter and astrophysical constraints.
The resulting equation of state can be incorporated in fluid-dynamical simulations of heavy-ion collisions and in simulations of neutron stars.
The simulations access the high density and moderate temperature
region of the QCD phase diagram, which can only be constraint by heavy ion collisions and observations of neutron star mergers.\\

The authors thank HIC for FAIR, HGS-HIRe for FAIR, the BMBF and DFG for support. In addition JS acknowledges support from the WGG-F\"orderverein and the F\"uck-Stiftung.





\bibliographystyle{elsarticle-num}
\bibliography{<your-bib-database>}



\end{document}